\documentclass[prb,twocolumn,showpacs,citeautoscript]{revtex4}

\usepackage{graphicx,epsfig}
\usepackage{bm}
\usepackage{multirow}
\usepackage{amsmath,amssymb,amsfonts}
\begin{document}

\title{Quantum phase transitions, frustration and the
       Fermi surface in the Kondo lattice model}
\smallskip

\author{Eitan Eidelstein, S. Moukouri
        and Avraham Schiller}

\affiliation{Racah Institute of Physics, The Hebrew
             University, Jerusalem 91904, Israel}

\begin{abstract}
The quantum phase transition from a spin-Peierls phase with
a small Fermi surface to a paramagnetic Luttinger-liquid
phase with a large Fermi surface is studied in the
framework of a one-dimensional Kondo-Heisenberg model
that consists of an electron gas away from half filling,
coupled to a spin-1/2 chain by Kondo interactions. The
Kondo spins are further coupled to each other with
isotropic nearest-neighbor and next-nearest-neighbor
antiferromagnetic Heisenberg interactions which are
tuned to the Majumdar-Ghosh point. Focusing on
three-eighths filling and using the density-matrix
renormalization-group (DMRG) method, we show
that the zero-temperature transition between
the phases with small and large Fermi momenta appears
continuous, and involves a new intermediate phase where
the Fermi surface is not well defined. The intermediate
phase is spin gapped and has Kondo-spin correlations
that show incommensurate modulations. Our results appear
incompatible with the local picture for the quantum phase
transition in heavy fermion compounds, which predicts an
abrupt change in the size of the Fermi momentum.
\end{abstract}



\maketitle


\section{Introduction}
\label{sec:Introduction}

The Kondo lattice model (KLM) was historically introduced
to describe the competition between singlet formation and
magnetic ordering in heavy fermion systems~\cite{Doniach}.
In heavy fermion materials, localized $f$-shell electrons
hybridize with itinerant electrons. Depending on whether
the $f$ electrons participate in the formation of the
Fermi surface (FS) or not, the latter may be large or
small~\cite{Hewson}. The Fermi momentum $k_{F}^{L}$ is
large in the sense that the FS encloses a volume
that counts both the number of conduction electrons
and local moments~\cite{Hewson,Oshikawa}. This is in
contrast to a small Fermi momentum $k_{F}^{S}$, 
whose FS encloses a volume that counts the number
of conduction electrons only.

An earlier issue debated on the KLM was whether the
formation of a large FS is consistent with Luttinger's
theorem~\cite{Luttinger}. In other terms, the question
was whether the KLM can account for a large FS given
that the $f$ electrons are represented by their spin
degrees of freedom only. Investigations of the
one-dimensional (1D) KLM through
numerical~\cite{Tsunetsugu-Sigrist-Ueda,
Shibata-Ueda-Nishino-Ishii,Shibata-Tsvelik-Ueda,
Moukouri-Caron,Sikkema-Affleck-White}
and analytical~\cite{Oshikawa,Yamanaka-Oshikawa-Affleck}
approaches have yielded rather consistent evidence for
a ground state with a large FS in the paramagnetic
regions of the model. Yet recent numerical calculations
have suggested the existence of a second phase where
the FS is small~\cite{Xavier-Novais-Miranda}. Such a
phase will necessary have a broken-symmetry ground state,
otherwise it would be inconsistent with Luttinger's
theorem.~\cite{Oshikawa,Yamanaka-Oshikawa-Affleck}
For reviews on the 1D KLM, we refer the reader to
Refs.~~\onlinecite{Tsunetsugu-Sigrist-Ueda}
and~~\onlinecite{Gulacsi}.

In the last decade, the question of the size of the FS
in the KLM has gained renewed interest in connection
with quantum criticality and the related non-Fermi-liquid
phases of heavy fermion materials. The local picture for
the quantum phase transition (QPT) in these compounds
predicts that the size of the FS would change abruptly at
the quantum critical point~\cite{CPR-01,Si-Zhu-Grempel}. 
The composite quasiparticles forming the large FS are
projected to breakdown as the system is driven across the
critical point, leaving behind a small Fermi volume that
counts the number of conduction electrons only. It is
still unclear how such a sudden change of the FS is
consistent with a second-order phase transition. 

One-dimensional models, for which powerful methods
of solution are available, are currently the
primary tool for gaining reliable information about
QPT in the KLM. However, magnetic orderings that break
spin-rotational symmetry are prohibited in 1D.
Hence, it is necessary to study the QPT between the
paramagnetic phase with a large FS and alternative
phases of the Ising or spin-Peierls
type~\cite{Pivovarov-Si}. If the $f$ electrons are in an
Ising or a dimerized phase, they would remain decoupled
from the conduction electrons at low energies also in
the presence of a sufficiently small Kondo coupling
as compared to the gap. Consequently, the FS would be
small. A QPT toward a ground state with a large FS
would occur upon increasing the Kondo coupling. This
latter phase is presumably a Luttinger liquid (LL)
or a spin-gapped phase. 

In this paper, we study the evolution of the ground
state of a 1D KLM from a spin-Peierls phase with a
small Fermi momentum $k_F^S$ to a LL phase with a
large Fermi momentum $k_F^L$. To this end, we augment
the conventional KLM with isotropic nearest-neighbor
and next-nearest-neighbor spin-exchange interactions
among the Kondo spins, and tune them to the
Majumdar-Ghosh point~\cite{Majumdar-Ghosh}. The
inclusion of next-nearest-neighbor coupling is a
crucial ingredient of our study, as it supports a
broken-symmetry ground state with a small FS for
small Kondo couplings.

Focusing on three-eighths filling and using the
density-matrix renormalization group method
(DMRG)~\cite{White-92}, we find a zero-temperature
transition that is more complex than the predictions of
the local critical theory. In particular, we identify
an intermediate spin-gapped phase in between the
spin-Peierls and LL phases where the Fermi momentum
cannot be defined. Instead, the electron momentum
distribution function $n(k)$ displays a shallow peak
at a new characteristic momentum $k^*$ that lies in
between $k_F^S$ and $k_F^L$, and which shifts toward
$k_F^L$ on going from the spin-Peierls to the LL
phase.  Concomitantly, there is a maximum at $2k^*$
in the magnetic structure factor $S(k)$ of the Kondo
spins. We show that this behavior can be understood
as a consequence of the magnetic frustration induced
by competing spin-exchange couplings generated by
the Ruderman-Kittel-Kasuya-Yosida (RKKY) interaction.
In contrast to $n(k)$ and $S(k)$, neither the Fourier
transform of the local conduction-electron density
$n_r(k)$ nor the density-density correlation function
$C(k)$ show any special signatures related to $k^*$.
Rather, the transition from the spin-Peierls to the
spin-gapped phase is manifest in $C(k)$ by the
smearing of a cusp at $2k_F^S$ and the emergence of
a peak at $2k_F^L$. Our results appear
incompatible with the local picture for the QPT in
heavy fermion compounds.
 
The remainder of the paper is organized as follows.
In Sec.~\ref{sec:model} we present the KLM under
study, along with details of our DMRG code. A
comprehensive set of results for the dimer order
parameter, the spin velocity, the electron
momentum-distribution function, the magnetic structure
factor of the Kondo spins, and various density correlations
are described and analyzed in Sec.~\ref{sec:results}.
We conclude in Sec.~\ref{sec:Conclusions} with a
discussion of the resulting phase diagram and its
relevance to heavy fermion compounds.


\section{The model }
\label{sec:model}

In this paper we study the following KLM
\begin{align}
H = & -t \sum_{i=1,\sigma}^{L-1}
         \left \{
                  c^{\dagger}_{i,\sigma}c_{i+1,\sigma}
                  + {\rm H.c.}
         \right\}
      + J_K \sum_{i=1}^{L}
            \vec{S}_{i} \cdot \vec{\tau}_{i}
\nonumber \\
    & + J_{H_1} \sum_{i=1}^{L-1}
                \vec{S}_{i} \cdot \vec{S}_{i+1}
      + J_{H_2} \sum_{i=1}^{L-2}
                \vec{S}_{i} \cdot \vec{S}_{i+2} ,
\label{hamiltonian}
\end{align}
describing a 1D tight-binding conduction band with
the hopping term $t$, interacting via an on-site
spin-exchange (Kondo) interaction $J_K$ with an array
of localized spins. Here, $c^{\dagger}_{i,\sigma}$
creates a conduction electron with spin-projection
$\sigma$ at site $i$, $\vec{S}_{i}$ represents the
localized spin at site $i$, and
$\vec{\tau}_{i} = \frac{1}{2} \sum_{\sigma,\sigma'}
c^{\dagger}_{i,\sigma}c_{i,\sigma'}
(\vec{\sigma})_{\sigma,\sigma'}$ is the
conduction-electron spin density at that site. The
conduction electrons and spins reside on an $L$-site
lattice with open boundary conditions (OBC).
In addition to the Kondo interaction, the localized spins
interact among themselves via the nearest-neighbor and
next-nearest-neighbor Heisenberg spin-exchange terms
$J_{H_1}$ and $J_{H_2}$, respectively. To avoid the
onset of ferromagnetism~\cite{Moukouri-Caron} we set
$J_{H_1}=t/2$, while $J_{H_2} = J_{H_1}/2$ is tuned
to the well-known Majumdar-Ghosh point,~\cite{Majumdar-Ghosh} 
whose corresponding ground state of the decoupled spin
chain is a perfect dimerized state (for even $L$).

As emphasized above, the inclusion of
$J_{H_2} = J_{H_1}/2 > 0$ is a crucial difference
from previous DMRG studies of the
KLM~\cite{Shibata-Ueda-Nishino-Ishii,Shibata-Tsvelik-Ueda,
Moukouri-Caron,Sikkema-Affleck-White,Xavier-Novais-Miranda,
Xavier-Pereira-Miranda-Affleck,Hotta-Shibata,
Xavier-Miranda}. This additional frustration opens
a gap in the spectrum of the isolated spin
chain~\cite{comment-on-gap}, enabling the study of
the transition from the broken-symmetry dimerized
phase with a small FS to the LL phase with a large
FS. From a technical standpoint, the dimerization gap
significantly reduces the numerical effort that is needed
to obtain reliable results as compared to the case where
$J_{H_1} = J_{H_2}=0$~\cite{Xavier-Pereira-Miranda-Affleck,
Hotta-Shibata,Xavier-Miranda}, due to the short-range
spin-spin correlations that develop.

We computed the ground state of the Hamiltonian of
Eq.~(\ref{hamiltonian}) using the DMRG method with
OBC. We retained between $256$ and $512$ states in
the two external blocks, keeping track of the total
number of electrons and the $z$ component of the
total spin projection $S_z$ as good quantum
numbers. The maximal truncation error was in the
order of $10^{-4}$ when $256$ states where kept 
and in the order of $10^{-5}$ when $512$ states where
kept. We studied different lattice sizes up to $L=64$
sites with $J_K$ varied in the range $0 \le J_K/t \le 16$.
For concreteness we set the conduction-electron filling
equal to $n = 0.75$, which is close to but off half
filling, and is rather convenient to tackle numerically.
We briefly comment on other filling factors at the end
of the paper. All results presented below are restricted
to zero temperature. The lattice size is $L = 64$
unless stated otherwise.


\section{Results}
\label{sec:results}

It is instructive to consider first the limits of small
and large $J_K$, where different phases are expected.
When $J_K=0$, the conduction electrons and spins are
decoupled, forming independent chains. The spin chain,
being tuned to the dimerized Majumdar-Ghosh phase, is
gapped due to the breaking of translational symmetry.
The electron chain is gapless in both the spin and
charge sectors, as is the overall system. Due to the
gap in the spectrum of the isolated spin chain, the
Majumdar-Ghosh phase is expected to be stable against
the inclusion of a small $J_K$.
 
\begin{figure}[tb]
\centerline{
\includegraphics[width=75mm]{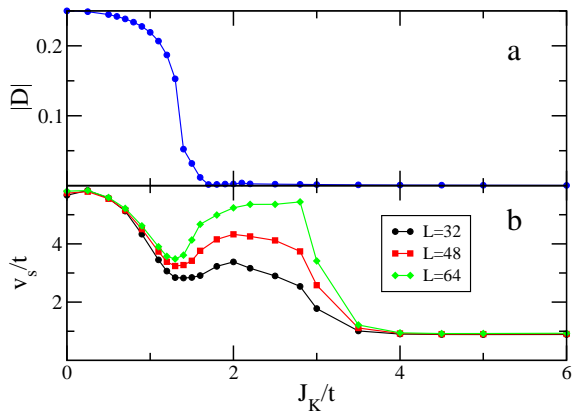}
}\vspace{0pt}
\caption{(Color online)
         (a) The dimer order parameter $|D|$ of
         Eq.~(\ref{dimers}), plotted as a function of
         $J_K/t$ for $n=0.75$ and zero temperature. For
         $J_K = 0$, the system is in a perfect dimer
         state, corresponding to $|D| = 0.25$. As the
         Kondo interaction $J_K$ is switched on the
         dimerization progressively decays until it
         vanishes for $J_K/t \agt 1.7$. Note the
         particularly sharp slope around
         $J_K/t = 1.3$--$1.4$.
         (b) The spin velocity $v_s$ of Eq.~(\ref{spinVel})
         as a function of $J_K/t$, plotted for different
         lattice sizes $L$ and $n = 0.75$. Different
         qualitative behaviors are observed for weak,
         intermediate, and strong couplings. In
         particular, a spin gap opens in the
         intermediate-coupling regime
         $1.3 \alt J_K/t \alt 4$.}
\label{Fig:spinVel_dimers}
\end{figure}

In the opposite limit $J_K \gg t,J_{H_1}$, the
conduction electrons and spins bind to form localized
singlets. At temperatures below $J_K$ there is no
thermal energy to break the Kondo singlets, which
can be viewed as holes in the underlying Kondo spin
texture. Hence, by analogy with the case where
$J_{H_2} = 0$~\cite{Sikkema-Affleck-White}, the system
is described by the $t-J_1-J_2$ model for holes, with
$t \to t/2$, $J_1 = J_{H_1}$, and $J_2 = J_{H_2}$.
Note that the original electronic filling factor $n$
(assumed to be smaller than $1$)
is converted by this mapping to the hole filling factor
$n_{\rm hole} = n$~\cite{comment-on-convention}.
The corresponding ground state of the
$t-J_1-J_2$ model is expected to be a paramagnetic LL
for $n_{\rm hole} = 0.75$, which differs in
symmetry from the dimerized phase at small $J_K$. Thus,
a QPT should occur upon increasing $J_K$. Below we
confirm this scenario and thoroughly discuss the
nature of the phase transition.


\subsection{Dimer order parameter}

The first quantity we study is the dimerization order
parameter $D$, defined as 
\begin{equation}
D = \frac{2}{3L}\sum_{i = 1}^{L-1}
            \langle
                    \vec{S}_{i} \cdot \vec{S}_{i+1}
            \rangle (-1)^i .
\label{dimers}
\end{equation}
Here the alternating $(-1)^i$ factor comes to distinguish
the dimerized phase from a translational-invariant state.  
Figure~\ref{Fig:spinVel_dimers}(a) depicts the evolution
of $|D|$ with increasing $J_K$. For $J_K = 0$ there is
perfect dimerization, corresponding to $|D| = 0.25$.
This value of $|D|$ stems from the fact that each spin
$\vec{S}_{i}$ forms a perfect singlet with one of its
neighbors and is uncorrelated with its other neighbor.
Consequently, $\langle \vec{S}_{i} \cdot \vec{S}_{i+1} \rangle$
equals $-3/4$ ($0$) for odd (even) $i$. With increasing
$J_K > 0$, the dimerization order parameter decreases
first gradually and then sharply around $J_K/t = 1.3-1.4$.
The sharp slope in the latter regime suggests a rather
rapid change in the nature of the ground state. Eventually
$|D|$ vanishes above $J_K/t \approx 1.7$, indicating
the loss of any remnant of the dimer state that forms
at small $J_K$. It should be noted that $D$ shows no
significant size dependence due to the short-range
spin-spin correlations that are involved.


\subsection{Spin velocity}

\begin{figure}[tb]
\centerline{
\includegraphics[width=75mm]{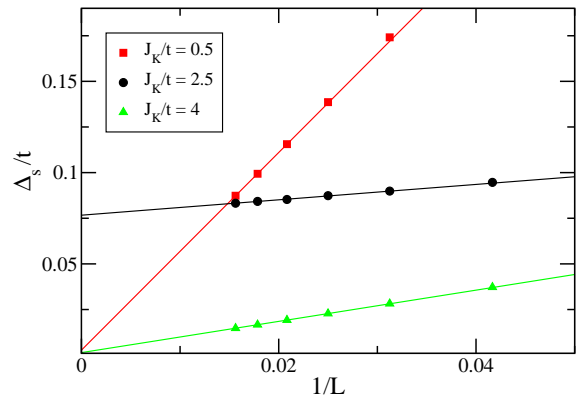}
}\vspace{0pt}
\caption{(Color online)
         The spin gap $\Delta_s(L)$ as a function of
         $1/L$, plotted for $n = 0.75$ and $J_K/t =
         0.5$, $2.5$, and $4$. The solid lines are
         linear extrapolations to $L \to \infty$. In
         contrast to $J_K/t = 0.5$ and $4$, where
         $\Delta_s(L)$ extrapolates to zero, a finite
         spin gap $\Delta_s(L \to \infty) \simeq 0.077 t$
         is found for $J_K/t = 2.5$~\cite{comment-on-D_s}.}
\label{Fig:spinGap}
\end{figure}

When $J_K/t \agt 1.7$ translational symmetry is restored,
as signaled by the vanishing of $D$. In order to better
understand the nature of this new phase we computed the
spin velocity $v_s$, defined as
\begin{equation}
	v_s (L)= \Delta_s(L) L .
\label{spinVel}
\end{equation}
Here $\Delta_s(L)$ is the elementary singlet-triplet
excitation energy for a system of size $L$.
Figure~\ref{Fig:spinVel_dimers}(b) shows $v_s(L)$ for
different lattice sizes as a function of $J_K$. When
$J_K/t \alt 1.3$, $v_s(L)$ strongly depends on $J_K$,
reflecting the progressive formation of a composite
quasiparticle made up of the local spins and the
itinerant electrons. By contrast, $v_s(L)$ is almost
independent of both $J_K$ and $L$ when $4 \alt J_K/t$.
This behavior can be understood from the fact that the
system is rather well described in this regime by the
$t-J_1-J_2$ model discussed above, which forms a LL
when $n_{\rm hole} = 0.75$. In the intermediate range,
$1.3 \alt J_K/t \alt 4$, $v_s(L)$ depends strongly
on both $J_K$ and $L$, indicating the emergence of
a nonzero spin gap $\Delta_s$ for $L \to \infty$.

To support this interpretation we have plotted
$\Delta_s(L)$ vs. $L$ in Fig.~\ref{Fig:spinGap}, for
three representative values of $J_K/t$. For both small
and large $J_K$ (represented by $J_K/t = 0.5$ and $4$,
respectively), $\Delta_s(L)$ extrapolates nicely to
zero as $L \to \infty$, indicative of a gapless
state. However, for the intermediate value of
$J_K/t=2.5$, $\Delta_s(L)$ extrapolates to the
finite spin gap $\Delta_s/t \approx 0.077$ as
$L \to \infty$~\cite{comment-on-D_s}. Such a
global spin gap is neither consistent with a
spin-dimerized phase nor with a LL phase. A
similar spin gap was found throughout the range
$1.3 \alt J_K/t \alt 4$, though the precise
boundaries of this new phase are somewhat difficult
to pin down~\cite{comment-on-phase-boundaries}.


\subsection{Electron momentum-distribution function}

Next we address the size of the FS and its evolution
upon going from $J_K = 0$ to $J_K/t = 16$, thereby
crossing the three different regimes of small,
intermediate and large $J_K$ signaled by $D$ and $v_s$
in Figs.~\ref{Fig:spinVel_dimers} and \ref{Fig:spinGap}.
As pointed out earlier in the introduction, previous
computations~\cite{Moukouri-Caron} on the KLM with
$J_{H_2} = 0$ concluded that a large FS forms in the
absence of symmetry breaking. Due to the gap in the
spectrum of the dimerized Majumdar-Ghosh phase, the
conduction electrons and spins remain decoupled at
low energies even in the presence of a small $J_K$,
thus forming a small FS. On the other hand, a large
FS should be recovered when $J_K \gg t, J_{H_1}$,
as the resulting behavior should basically reproduce
that of $J_{H_2} = 0$. 

\begin{figure}[tb]
\centerline{
\includegraphics[width=75mm]{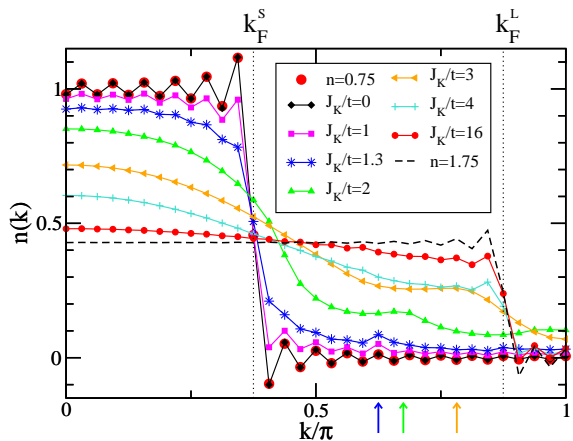}
}\vspace{0pt}
\caption{(Color online)
         The electron momentum-distribution function
         of Eq.~(\ref{nOfk}), for $n = 0.75$ and
         different Kondo couplings $J_K/t = 0,1$,
         $1.3$, $2$, $3$, $4$ and $16$. When $J_K = 0$,
         the conduction electrons form a free band,
         whose exact momentum-distribution function is
         depicted by the red dots. Accordingly, there
         is a sharp step in the momentum-distribution
         function at the (small) Fermi momentum
         $k_{F}^S = \pi n/2 = 0.375 \pi$. The step at
         $k_{F}^S$ persists for small Kondo interactions,
         $J_K/t \alt 1.3$. In the opposite limit of large
         $J_K$ (represented by $J_K/t = 16$), there
         is a new step at the large Fermi momentum
         $k_{F}^L = \pi (n + 1)/2 = 0.875 \pi$.
         The corresponding momentum-distribution function
         compares rather well with that of a noninteracting
         tight-binding chain of equal length and the
         filling $n' = 1+n = 1.75$, scaled down by a
         factor of $n/n' = 3/7$ (black dashed line).
         In contrast to the limits of small and large
         $J_K$, there is no clear sign of a FS for the
         intermediate couplings $J_K/t = 2$ and $3$.}
\label{Fig:nOfk_0.75}
\end{figure}

In order to study the transition between these vastly
different Fermi momenta, we computed the electron
momentum-distribution function, defined as
\begin{equation}
n(k) = \sum_{i = 1}^{L}
       \sum_{\sigma}
            \langle
                    c^{\dagger}_{i, \sigma} c_{L/2, \sigma}
            \rangle \cos[k(i-L/2)].
\label{nOfk}
\end{equation}
This definition of $n(k)$ differs from the conventional
one $n(k) = \sum_{\sigma} \langle c^{\dagger}_{k, \sigma}
c_{k, \sigma} \rangle$ due to the OBC used. Nevertheless,
it contains similar information on the FS, as we show
below. In particular, the two definitions must coincide
in the thermodynamic limit $L \to \infty$, provided that
translational symmetry is not broken. To see this we note
that the correlator $ \langle c^{\dagger}_{i, \sigma}
c_{L/2, \sigma} \rangle$ with fixed $i - L/2$ becomes
independent of the boundary conditions for
$L \to \infty$. Converting to periodic boundary
conditions and in the presence of translational invariance,
Eq.~(\ref{nOfk}) reduces then to $\frac{1}{2} \sum_{\sigma}
\left[ \langle c^{\dagger}_{k,\sigma} c_{k, \sigma} \rangle +
\langle c^{\dagger}_{-k, \sigma} c_{-k, \sigma} \rangle\right]$,
which is nothing but the conventional momentum-distribution
function. Here we made use of the fact that
$\langle c^{\dagger}_{k, \sigma} c_{k, \sigma} \rangle$ and
$\langle c^{\dagger}_{-k, \sigma} c_{-k, \sigma} \rangle$
are identical by virtue of inversion symmetry.

Figure~\ref{Fig:nOfk_0.75} shows $n(k)$ for different
Kondo interactions at the fixed filling factor $n=0.75$.
For $J_K=0$, one can compute $n(k)$ exactly from
the single-particle eigenstates of the decoupled
conduction-electron chain. The exact results, depicted
by the red circles, essentially coincide with our DMRG
data, serving as a critical check for the accuracy
of our code. As expected, there is a sharp step in
$n(k)$ at the small Fermi momentum
$k_{F}^S = \pi n/2 = 0.375 \pi$ of the free band.
Note that $n(k)$ oscillates as a function of $k$,
and can become both negative and may exceed one.
These finite-size effects are eliminated in the
thermodynamic limit $L \to \infty$. We emphasize,
however, that the definition of Eq.~(\ref{nOfk})
does not require that $0 \leq n(k) \leq 1$.

As soon as the Kondo interaction is switched on, the
system gradually looses the sharpness of the FS, as
expected of an interacting 1D system. Nevertheless, a
clear Fermi momentum can still be observed at $k_{F}^S$
for $J_K/t \alt 1.3$, i.e., in the range where a
sizeable dimerized order persists (see
Fig.~\ref{Fig:spinVel_dimers}). In the strong-coupling
limit $4 \alt J_K/t$ (note that $4t$ is the free
conduction-electron band-width), a new well-defined Fermi
momentum appears, this time at $k_F^L = \pi (n + 1)/2
= 0.875\pi$. This value of $k_F$ corresponds to a FS
which encloses a volume that counts both the number of
conduction electrons (filling factor of $n = 0.75$)
and the number of local moments (``filling factor''
of $n = 1$). For comparison, the black dashed line
represents the exact momentum-distribution function
for a noninteracting tight-binding chain of equal
length and the filling $n' = 1+n = 1.75$, scaled
down by a factor of $n/n' = 3/7$. This latter
renormalization reflects the fact that the actual
conduction-electron filling in our system [corresponding
to the integrated weight of $n(k)$] is $n$ rather than
$n' = 1+n$. The agreement is quite surprising.

In contrast to the weak- and strong-coupling regimes,
there is no well-defined Fermi momentum for the
intermediate couplings $J_K/t = 2$ and $3$. Instead,
the sharp structure at $k_F^S$ is rapidly smoothed
and suppressed, and a new feature appears at a
characteristic momentum $k^*$ located in between
$k_{F}^{S}$ and $k_{F}^{L}$. This new feature,
whose position is marked by the arrows in
Fig.~\ref{Fig:nOfk_0.75}, first appears for $J_K/t = 1.3$
as a small and shallow peak at $k^* = 0.625\pi$. It
continuously shifts toward $k_F^L$ upon increasing $J_K$,
until it coincides with the new Fermi momentum $k_F^L$
for $J_K/t = 4$. As seen from the Friedel oscillations
and the density-density correlation
function presented below, $k^*$ is not associated
with a new Fermi momentum. Rather, it reflects the
spin-spin correlations that develop in this range due
to the combination of $J_{H_1}$, $J_{H_2}$, and the
RKKY interaction mediated by the conduction electrons.
In other words, $k^*$ stems from the back action of
the presumably frustrated spins on the conduction
electrons.


\subsection{Magnetic structure factor}

\begin{figure}[tb]
\centerline{
\includegraphics[width=75mm]{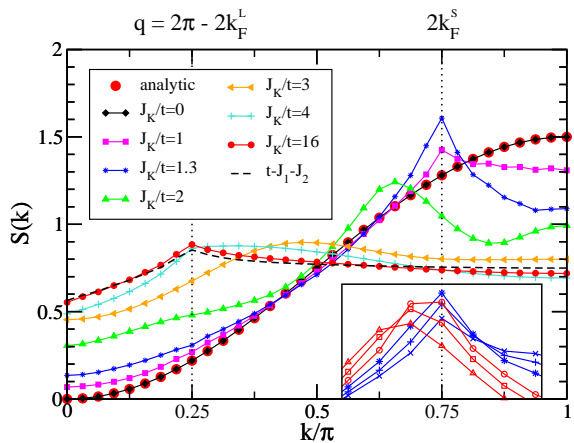}
}\vspace{0pt}
\caption{(Color online)
         The magnetic structure factor of Eq.~(\ref{sOfk}),
         plotted for $n = 0.75$ and the same values of
         $J_K$ as in Fig.~\ref{Fig:nOfk_0.75}. For
         $J_K = 0$, the Kondo spins are locked in a
         perfect dimer state. The exact magnetic structure
         factor in this case (red dots) has a broad peak
         at $k = \pi$. Once $J_K$ is switched on, a sharp
         cusp gradually develops at $2k_{F}^{S} = 0.75\pi$,
         reaching a maximum for $J_K/t \approx 1.3$. Upon
         further increasing $J_K$, the cusp smoothens
         and shifts continuously all the way to
         $q = 2\pi - 2k_{F}^{L} = 0.25\pi$, where a new
         cusp forms at large $J_K$. In the limit of large
         $J_K$, represented by $J_K/t = 16$, $S(k)$ well
         matches the structure factor of the corresponding
         $t-J_{1}-J_{2}$ model, which is shown for
         comparison by the black dashed line.
         Inset: Evolution of the cusp at $2k_F^S$ as it
         first begins to move. Here crosses, pluses,
         stars, circles, squares, and triangles
         correspond to $J_K/t = 1.1$, $1.2$,
         $1.3$, $1.4$, $1.5$, and $1.6$, respectively. For
         $J_K/t \leq 1.3$ (blue lines), the cusp is pinned
         to $2k_F^S$, growing in magnitude with increasing
         $J_K$. For $1.4 \leq J_K/t$ (red lines), the cusp
         is shifted to incommensurate modulations.}
\label{Fig:sOfk_0.75}
\end{figure}

The structure that $n(k)$ develops at $k^*$ in the
intermediate phase is rather small. We now show that
a much more pronounced feature is found at $2k^*$
in the magnetic structure factor of the Kondo spins,
defined as
\begin{equation}
S(k) = \sum_{i = 1}^{L}
            \langle
                    \vec{S}_{i} \cdot \vec{S}_{L/2}
            \rangle \cos[k(i-L/2)].
\label{sOfk}
\end{equation}
Similar to $n(k)$, Eq.~(\ref{sOfk}) is defined to match
the OBC used. It reduces in the thermodynamic limit and
in the absence of translational symmetry breaking to the
Fourier transform of the spin-spin correlation function
$\langle \vec{S}_{i} \cdot \vec{S}_{j} \rangle$. 

Figure~\ref{Fig:sOfk_0.75} shows the evolution of
$S(k)$ with increasing $J_K$. For $J_K=0$, when the
spin chain locks in a perfect dimerized ground state,
$S(k)$ equals $0.75[1-\cos(k)]$. This curve, depicted
by the red dots in Fig.~\ref{Fig:sOfk_0.75}, has a
maximum at $k=\pi$ and is well reproduced by our DMRG
data. In the opposite limit of a large $J_K$, $S(k)$
reduces to the magnetic structure factor of the
$t-J_1-J_2$ model, which has a cusp at
$q = 2\pi - 2k_F^L = \pi (1 - n) = 0.25\pi$ [note
that $q$ and $2k_F^L$ are equivalent by virtue of
$S(k) = S(2\pi - k)$, and will be regarded as
synonymous hereafter]. The transition between
these two limits proceeds as follows.

When $J_K$ is switched on, the spins couple to the
conduction electrons. As a result, $S(k)$ gradually
deforms from its perfect dimerized profile and develops
a sharp cusp at $2k_F^S = 0.75\pi$, as can be seen for
$J_K/t = 1$ and $1.3$. The evolution of this cusp is
tracked in the inset. With increasing $J_K$ the cusp
initially grows sharper, until reaching a maximum
for $J_K/t \approx 1.3$. Upon further increasing $J_K$
($1.4 \le J_K/t \le 4$), the cusp smoothens and
shifts continuously all the way to $q = 0.25\pi$.
Interestingly, the cusp at $2k_F^S$ first begins to shift
at the same value of $J_K$ where the dimerized order
parameter $D$ acquires its sharpest slope. Furthermore,
in the intermediate regime $1.4 \le J_K/t \alt 4$, the
associated peak in $S(k)$ occurs at $q^* = 2\pi - 2k^*$,
where $k^*$ is the position of the additional structure
found in $n(k)$. Hence the two features are intimately
related. Lastly we note that the local maximum that is
seen at $k = \pi$ for $J_K/t = 2$ is a finite-size effect
that decreases in magnitude with increasing system size.
This is in contrast to the incommensurate peaks at $q^*$
which depend only weakly on $L$.

\begin{figure}[tb]
\centerline{
\includegraphics[width=75mm]{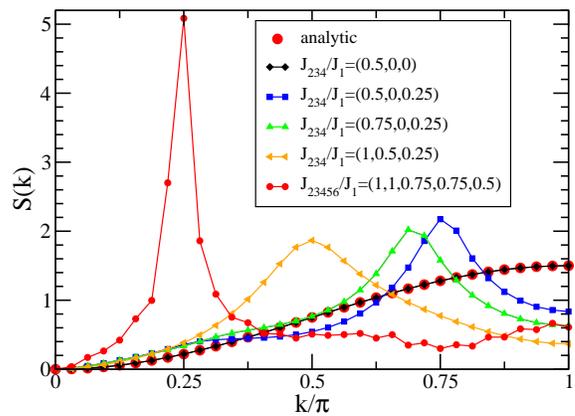}
}\vspace{0pt}
\caption{(Color online)
         The magnetic structure factor of the effective
         Heisenberg Hamiltonian with different
         spin-exchange couplings $J_1, J_2, \ldots,
         J_6$, up to a distance of $6$ lattice sites.
         Starting from the Majumdar-Ghosh point and
         gradually increasing the relative strengths
         of $J_2$ to $J_6$, we are able to shift the
         main modulation in $S(k)$ from $k = \pi$ to
         $2k_F^S = 0.75\pi$, and then all the way to
         $q = 0.25\pi$, in close analogy to the peak
         positions observed in Fig.~\ref{Fig:sOfk_0.75}.
         The incommensurate modulations that are seen
         in Fig.~\ref{Fig:sOfk_0.75} can thus be
         understood as due to the frustrated RKKY
         interaction, whose magnitude increases with
         $J_K$.}
\label{Fig:Heisenberg}
\end{figure}

In order to understand the origin of the incommensurate
modulations in $S(k)$, we studied an effective spin
Hamiltonian which comes to mimic the RKKY interactions
present in the KLM. Specifically, we consider a
frustrated Heisenberg model with different spin-exchange
interactions $J_1, J_2, \ldots, J_6$, up to a distance
of $6$ lattice sites. Such an effective Hamiltonian
with adequate couplings is expected to properly describe
the spin dynamics in the KLM up to moderately large
values of $J_K$. This description must eventually
break down for large $J_K$, when the conduction
electrons and spins tightly bind to form localized
singlets.

Figure~\ref{Fig:Heisenberg} shows the magnetic structure
factor $S(k)$ for the effective Heisenberg model. As can
be seen, we are able to generate incommensurate modulations
in $S(k)$ by varying the relative strengths of the
different spin-exchange interactions. In particular,
starting from the Majumdar-Ghosh point and gradually
increasing the relative strengths of $J_2$ to $J_6$,
we are able to shift the main modulation from $k = \pi$
(perfect dimers) to $2k_F^S = 0.75\pi$, and then all
the way to $q = 0.25\pi$. Note that the peak positions
in Fig.~\ref{Fig:Heisenberg} are quite similar to
those in Fig.~\ref{Fig:sOfk_0.75}, though their
profiles progressively deviate from those in
Fig.~\ref{Fig:sOfk_0.75} upon increasing the couplings.
This is to be expected from the approach to strong
coupling, where the KLM reduces to the $t-J_1-J_2$
model. These results indicate that the effective
RKKY interaction between the Kondo spins is
responsible for the incommensurate modulations
observed in Fig.~\ref{Fig:sOfk_0.75}.


\subsection{Density correlations and charge structure factor}

\begin{figure}[tb]
\centerline{
\includegraphics[width=75mm]{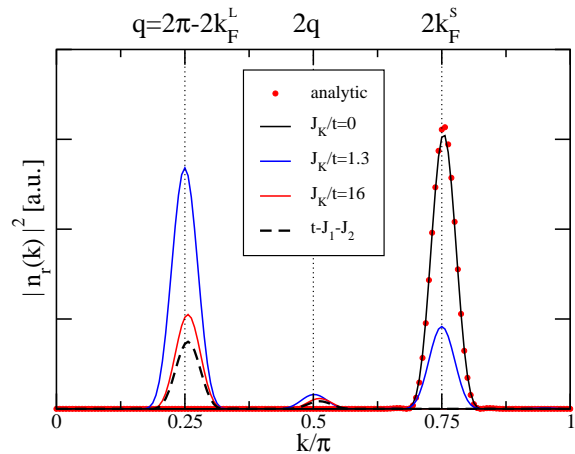}
}\vspace{0pt}
\caption{(Color online)
         The smoothed Fourier transform of the local
         density defined in Eq.~(\ref{nrOfk}), for
         $n = 0.75$ and different Kondo interactions.
         For $J_K = 0$, there are pronounced Friedel
         oscillations at the modulation wavelength
         $2 k_F^S = 0.75 \pi$, in good agreement with
         an exact evaluation of Eq.~(\ref{nrOfk}) using
         the single-particle eigen-modes of the isolated
         tight-binding chain (red circles). At large
         $J_K$, represented by $J_K/t = 16$, there
         are Friedel oscillations at both
         $q = 2\pi - 2k_{F}^{L} = 0.25\pi$ and
         $2q = 0.5\pi$, in accordance with $|n_r(k)|^2$
         for the corresponding $t-J_{1}-J_{2}$ model
         (black dashed line). Surprisingly, for
         $J_K/t = 1.3$ there are simultaneous
         modulations at $2k_F^S$, $q$, and $2q$, as
         if there were two coexisting Fermi momenta
         $k_F^S$ and $k_F^L$ (note that
         $2q = 2\pi - 4 k_F^S = 4\pi - 4 k_F^L$
         corresponds to both $4 k_F^S$ and $4 k_F^L$).}
\label{Fig:Friedel_oscillations_0.75}
\end{figure}

In order to further prove the spin origin of the feature
that $n(k)$ displays at $k^*$, we proceed to show results
on density correlations and the charge structure factor.
Figure~\ref{Fig:Friedel_oscillations_0.75} depicts
the smoothed Fourier transform of the local density
$n_{r}(k)$, defined as
\begin{equation}
n_{r}(k) = \sum_{i = 1}^{L}
           \left[
                  \langle \hat{n}_{i} - n \rangle
                  - \frac{I}{J}
           \right]
           W(i) \cos\!\left[
                             k\!\left(\frac{L+1}{2}-i\right)
                      \right].
\label{nrOfk}
\end{equation}
Here, $n = 0.75$ is the filling factor, $\hat{n}_{i}$
equals $\sum_{\sigma} \hat{n}_{i, \sigma}$ with
$\hat{n}_{i,\sigma} = c^\dagger_{i,\sigma}c_{i,\sigma}$,
and $W(i)$ is a smooth windowing function introduced
to avoid spurious edge
effects~\cite{White-Affleck-Scalapino,Vekic-White}.
The additional term $I/J$ with
$I = \sum_{i = 1}^{L} \langle n_{i} - n\rangle W(i)$
and $J = \sum_{i = 1}^{L}W(i)$ comes to correct for
the weighted average of $\hat{n}_{i}$, which slightly
deviates from $n$. Its role is to remove an artificial
feature near $k = 0$ introduced by $W(i)$.

The Fourier transform of the local density
oscillations is a measure of the Friedel oscillations
that develop due to the OBC used. It thus offers a
direct way to probe the Fermi momentum, which is
manifest as a peak at $2k_F$ and its higher
harmonics~\cite{White-Affleck-Scalapino,Vekic-White}.
Figure~\ref{Fig:Friedel_oscillations_0.75} shows the
Friedel oscillations for three values of $J_K/t = 0$,
$1.3$ and $16$. For $J_K = 0$ we observe a single peak
at $2k_F^S = 0.75\pi$, which well agrees with an
exact evaluation of Eq.~(\ref{nrOfk}) using the
single-particle eigen-modes of the decoupled
conduction-electron chain (red circles). In the
opposite limit of large $J_K$, there is one peak at
$q = 2\pi - 2k_F^L = 0.25\pi$ and another peak at
$2q = 0.5\pi$~\cite{comment-on-shifted-peaks}.
Both structures agree favorably with similar
calculations for the corresponding $t-J_{1}-J_{2}$
model~\cite{comment-on-t-j-j}, whose results are
shown by the black dashed line. Remarkably, for
the intermediate coupling $J_K/t = 1.3$ we find
modulations at all three momenta $2k_F^S$, $q$,
and $2q$, as if there were two coexisting Fermi
momenta~\cite{comment-on-6k_F^S} $k_F^S$ and $k_F^L$
(note that $2q$ corresponds both to $4 k_F^S$ and
$4 k_F^L$). These results are in stark contrast
to the electron momentum-distribution function
of Fig.~\ref{Fig:nOfk_0.75}, which shows only a
single step at $k_F^S$ for this value of $J_K$.

\begin{figure}[tb]
\centerline{
\includegraphics[width=75mm]{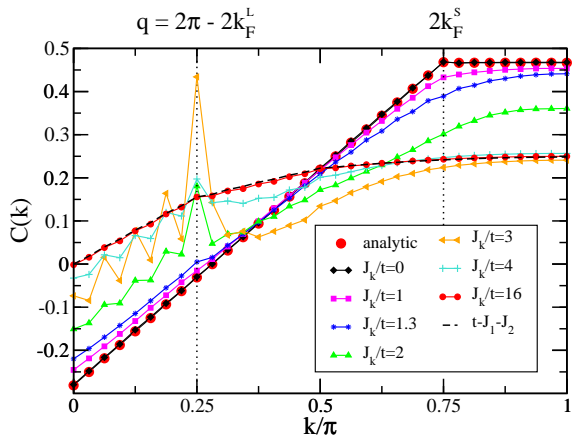}
}\vspace{0pt}
\caption{(Color online)
         Fourier transform of the density-density
         correlation function of Eq.~(\ref{ninjOfk}),
         for $n = 0.75$ and different Kondo
         interactions. For $J_K = 0$ there is a
         single sharp cusp at $2k_F^S = 0.75\pi$,
         in good agreement with the exact
         noninteracting result depicted by the red
         circles. For large $J_K$, the sharp structure
         at $2k_F^S$ is replaced with two very shallow
         cusps at $q = 2\pi - 2k_F^L = 0.25\pi$ and $2q$,
         in agreement with $C(k)$ for the corresponding
         $t-J_{1}-J_{2}$ model~\cite{comment-on-t-j-j}
         (black dashed line). At the intermediate
         couplings $J_K/t = 2$, $3$, and $4$ there is
         a sharp modulation at $q$, which degenerates
         into the shallow cusp of the $t-J_{1}-J_{2}$
         model upon increasing $J_K$.}
\label{Fig:ninjOfk_0.75}
\end{figure}

As a function of $J_K$, the peak at $2k_F^S$ persists
from $J_K = 0$ up to $J_K/t \approx 1.7$, which is the
point where the dimerized order parameter $|D|$ first
vanishes (see Fig.~\ref{Fig:spinVel_dimers}). Above
$J_K/t \approx 1.7$ there are no discernable signatures
left at $2k_F^S$. The peak at $q$ appears continuously
at small $J_K$, and persists all the way up to large
$J_K$. It evolves, however, in a nonmonotonous fashion,
first growing in magnitude before decreasing toward its
asymptotic $t-J_1-J_2$-model form. Interestingly, no
special signature appears at the incommensurate momentum
$q^* = 2\pi - 2k^*$, indicating that $k^*$ is unrelated
to the FS.
 
Finally, in Fig.~\ref{Fig:ninjOfk_0.75} we show
the Fourier transform of the conduction-electron
density-density correlation function, defined as
\begin{equation}
C(k) = \sum_{i = 1}^{L}
            \langle (n_{i} - n) (n_{L/2} - n) \rangle 
            \cos[k(i-L/2)] .
\label{ninjOfk}
\end{equation}
Similar to previous plots, the DMRG data well reproduce
the exact noninteracting result for $J_K = 0$, and are
consistent with the $t-J_{1}-J_{2}$ model for large
$J_K$. For $J_K = 0$ there is a single sharp cusp at
$2k_F^S = 0.75\pi$, which gradually smoothens as $J_K$
is switched on. For large $J_K$ (represented by
$J_K/t = 16$), there are two very shallow cusps at
$q$ and $2q$, corresponding to modulations at $2 k_F^L$
and $4 k_F^L$. Both limits of small and large $J_K$ are
compatible with the Friedel oscillations observed in
Fig.~\ref{Fig:Friedel_oscillations_0.75}.

A more complex structure is found for the intermediate
couplings $J_K/t=2$, $3$ and $4$. Here a sharp
modulation develops at $q$, accompanied by
additional wiggles at smaller values of $k$. We
believe that the latter wiggles are a finite-size
effect since they can be practically removed by using
a windowing function that smoothly falls off toward
the chain edges. By contrast, the primary peak at $q$
not only persists but actually grows in magnitude if
a windowing function is used. Similar to the Friedel
oscillations, the primary peak at $q$ evolves
nonmonotonically with increasing $J_K$, first
growing in magnitude before degenerating into the
shallow cusp of the $t-J_{1}-J_{2}$ model for
large $J_K$. The overall behavior of $C(k)$ in
this regime is consistent with a tendency toward
charge-density-wave or superconducting correlations
that may accompany the opening of a spin gap. Lastly
we note that $C(k)$ shows no discernable signature
related to $k^*$, reinforcing the spin origin of
this modulation wavelength.


\section{Discussion and conclusions}
\label{sec:Conclusions}

\begin{figure}[t]
\centerline{
\includegraphics[width=75mm]{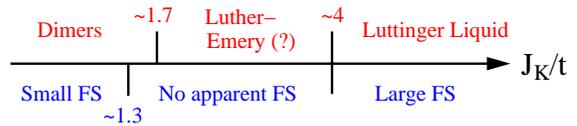}
}\vspace{0pt}
\caption{(Color online)
         Phase diagram of the KLM of Eq.~(\ref{hamiltonian}),
         for $J_{H_1} = 2J_{H_2} = t/2$ and $n = 0.75$.
         At weak coupling, $0 \leq J_K/t \alt 1.3$, the
         system is in a gapless dimer state with the
         small Fermi momentum $k_{F}^{S} = \pi n/2$. A
         spin-gapped phase sets in at intermediate
         coupling $1.3 \alt J_K/t \alt 4$, part of
         which retains nonzero dimer order (for
         $1.3 \alt J_K/t \alt 1.7$) and part of
         which where no dimer order is left (for
         $1.7 \alt J_K/t \alt 4$). The spin-gapped
         phase lacks a clear Fermi momentum. Finally,
         for $4 \alt J_K/t$ the system enters a
         paramagnetic LL phase with
         $k_{F}^{L} = \pi (n + 1)/2$ serving as the
         new Fermi momentum.}
\label{phases}
\end{figure}

We begin by summarizing the zero-temperature phase
diagram of the KLM of Eq.~(\ref{hamiltonian}), for
$J_{H_1} = 2J_{H_2} = t/2$ and $n = 0.75$.  For
weak Kondo couplings, $0 \leq J_K/t \alt 1.3$, the
system is in a gapless dimer state with the small Fermi
momentum $k_{F}^{S} = \pi n/2= 0.375 \pi$. At intermediate
Kondo couplings, $1.3 \alt J_K/t \alt 4$, a spin-gapped
phase sets in~\cite{comment-on-phase-boundaries}, part
of which retains nonzero dimer order $|D|$
(for $1.3 \alt J_K/t \alt 1.7$),
and part of which where no dimer order is left (for
$1.7 \alt J_K/t \alt 4$). In the latter regime the
system is presumably a Luther-Emery liquid. In contrast
to the dimer state at weak coupling, the spin-gapped
phase lacks a clear FS. There is no distinctive step
in the electron momentum-distribution function, and the
magnetic structure factor for the Kondo spins displays
incommensurate modulations at $2 k^{\ast}$, where
$k^{\ast}$ is a new characteristic momentum. The latter
wavelength is not associated with a new Fermi momentum,
as neither the Friedel oscillations encoded in $n_r(k)$
nor the density-density correlation function $C(k)$
show any discernable signatures related to
$k^{\ast}$. Rather, above $J_K/t \approx 1.7$
both $n_r(k)$ and $C(k)$ display pronounced
modulations related to $k_F^{L}$ alone, in
contrast to the regime $J_K/t \alt 1.7$ where
$n_r(k)$ has simultaneous modulations at
$2 k_{F}^{S}$ and $2 k_{F}^{L}$. Finally, for
$4 \alt J_K/t$ the system enters the paramagnetic
LL phase of the effective $t-J_1-J_2$ model, with
$k_{F}^{L}$ serving as the new Fermi momentum. A
sketch of the resulting phase diagram of the KLM
is displayed in Fig.~\ref{phases}. 

It should be emphasized that the opening of a spin gap
in the KLM is by itself not new. A spin-gapped phase
has long been established~\cite{Sikkema-Affleck-White}
for the Kondo-Heisenberg model with $n < 1$ and
$J_{H_2} = 0$, including recent indications for
quasi-long-range superconducting correlations at a
nonzero momentum~\cite{Berg-Fradkin-Kivelson}.
Here, however, the spin-gapped phase shows up as an
intermediate phase, separating two paramagnetic
phases with different Fermi momenta. This is quite
different from previous results for
$J_{H_2} = 0$,~\cite{Sikkema-Affleck-White} where
a single gapless phase was reported for values
of $J_K$ that exceeded the spin-gapped phase.

Our results should be compared to those of
Pivovarov and Si~\cite{Pivovarov-Si}, who used a
perturbative renormalization-group (RG) analysis
to study a more general form of the KLM, including
the effect of spin-exchange anisotropy in $J_{H_1}$
and $J_{H_2}$. For the SU(2) spin-symmetric interactions
considered here, the perturbative RG and DMRG
results differ in two respects:
(i) Perturbative RG predicts a region of coexistence
    between the weak-coupling spin-Peierls phase and
    the strong-coupling Kondo-singlet phase, whereas
    no such region is found by the DMRG. 
(ii) The strong-coupling phase is predicted by RG to
     have a spin gap, while our DMRG study finds a LL.
Indeed, we have verified by explicit calculations that
the $t-J_1-J_2$ Hamiltonian onto which the system maps
at strong coupling is gapless for $n_{\rm hole} = 0.75$,
thus forming a LL. Note, however, that a spin-gapped
phase is expected in the $t-J_1-J_2$ model as well
when tuned sufficiently close to half filling (i.e.,
for $n \ll 1$ in the KLM).

The transition studied in this work does not appear
to be consistent with the local picture for the QPT
in heavy fermion compounds~\cite{CPR-01,Si-Zhu-Grempel}.
We did not observe a sudden change in the size of the
FS, despite the fact that such a change would be more
favorable in 1D where the FS is reduced to just two
isolated points. The intermediate spin-gapped phase for 
$1.3 \alt J_K/t \alt 4$ seems to be a region where
the Fermi surface reconstructs from $k_F^S$ to $k_F^L$. 

It is not obvious to what extent can the QPT studied
in this work be compared to transitions involving
true magnetic order, as seen experimentally in heavy
fermions systems. For instance, the opening of a spin
gap in the intermediate phase appears to be related to
our choice of the Majumdar-Ghosh ground state for the
isolated spin chain. Nevertheless, frustration can
arise directly from the RKKY interaction itself, as
is known to occur at quarter
filling~\cite{Xavier-Pereira-Miranda-Affleck}. We have
studied the Hamiltonian of Eq.~(\ref{hamiltonian}) for
$n = 0.5$ both with and without explicit frustration
and found the same qualitative behavior. The electron
momentum-distribution function $n(k)$ displayed the
same general behavior as for $n = 0.75$, including an
intermediate region with no apparent Fermi momentum.
At the same time, we did not find any spin gap in
the intermediate regime, nor did we identify any new
characteristic momentum $k^{\ast}$ associated with a
shifting structure in either $n(k)$ or $S(k)$. The
absence of $k^{\ast}$ for $n = 0.5$ likely stems
from the fact that $2k_F^S$ and $2k_F^L$ are
indistinguishable for this particular filling factor.

This naturally raises the question of how generic
is the transition observed for $n = 0.75$. Is it
representative of other filling factors in
Eq.~(\ref{hamiltonian}) or does the QPT vary
qualitatively as a function of $n$? Preliminary
results for $n = 0.25$ and $n = 0.875$ suggest
the following~\cite{comment-on-ph-symmetry}.
For all filling factors studied there is an
intermediate region where $k_F$ cannot be defined.
This aspect, as well as the overall behavior of the
momentum-distribution function, appears to be generic.
However, details of the intermediate region do depend
on $n$. As stated above, we did not find any spin gap
for $n = 0.5$, in contrast to the filling factors
$n = 0.75$ and $0.875$ which are both spin gapped and
in qualitative agreement with each other. The picture
for $n = 0.25$ seems to be more complex, and may
potentially involve more than one intermediate phase.
This possibility, currently under study, may suggest a
qualitative difference between the regimes $n < 0.5$
and $0.5 < n < 1$.

To conclude, we conducted an exhaustive investigation
of the transition from a spin-Peierls phase with a
small Fermi momentum to a LL phase with a large Fermi
momentum in a 1D KLM. Our findings indicate a rather
complex transition that exceeds the predictions of
local criticality. It remains to be seen which of our
results extend to other electronic fillings and other
variants of the 1D KLM, let alone to higher dimensions.
Further investigations of this fascinating issue are
clearly in order.

\begin{acknowledgments}
We are grateful to Dror Orgad and Efrat Shimshoni for
illuminating discussions. We are particularly thankful
to Efrat Shimshoni for stimulating our interest in
this problem. This work was supported in
part by a Shapira fellowship of the Israeli Ministry
of Immigrant Absorption (S.M.), and by the Israel
Science Foundation through grant no. 1524/07.
\end{acknowledgments}


\begin{thebibliography}{30}
\bibitem{comment-on-gap}
         An infinite system possesses two degenerate
         ground states, each displaced by one lattice
         site with respect to the other, and a finite
         energy gap to the first excitation energy.
         This ground-state degeneracy is lifted for
         a finite system with OBC, yet is quickly
         restored as $L \to \infty$. In contrast, the
         energy gap to the second excited state saturates
         at a nonzero constant as $L \to \infty$.
\bibitem{comment-on-convention}
         We use the prevailing convention whereby the
         hole filling factor is measured with respect
         to a half-filled band. In other words,
         $n_{\rm hole}$ measures the fraction of
         empty sites in the $t-J_{1}-J_{2}$ model.
\bibitem{comment-on-D_s}
         The data displayed in Fig.~\ref{Fig:spinGap}
         was obtained keeping $512$ states in each of
         the external blocks. Only a small variation of
         order 1\% was found in $\Delta_s(L \to \infty)$
         for $J_K/t = 2.5$ when $256$ states were kept.
\bibitem{comment-on-phase-boundaries}
         As can be seen in Fig.~\ref{Fig:spinGap},
         there is no spin gap for $J_K/t = 4$. Indeed,
         as a function of $J_K$ the gap appears to close
         somewhere between $J_K/t = 3$ and $J_K/t = 4$,
         which defines the location of the upper phase
         boundary. The position of the lower phase
         boundary is harder to pinpoint. We estimate it
         to lie close to $J_K/t = 1.3$, where both the
         dimer order parameter $|D|$ and the spin
         velocity $v_s$ show significant changes in
         behavior.
\bibitem{comment-on-shifted-peaks}
         Due to the introduction of a windowing function,
         there is a slight shift in the positions of
         the peaks away from $q$ and $2q$. We have
         confirmed that these shifts depend on the
         choice of $W(i)$ and are not inherent to
         the model.
\bibitem{comment-on-t-j-j}
         Strictly speaking, the mapping onto the
         $t-J_{1}-J_{2}$ model requires that $n_r(k)$
         and $C(k)$ be computed using the local hole
         (i.e., empty-site) densities rather than the
         electronic densities. However, $|n_r(k)|$ and
         $C(k)$ are identical for holes and electrons
         in the $t-J_{1}-J_{2}$ model due to the
         omission of doubly occupied states.
\bibitem{comment-on-6k_F^S}
         Note that the peak at $0.25 \pi$ can also be
         interpreted as corresponding to $6 k_F^S$.
         Nevertheless, we find such a pronounced
         modulation at $6 k_F$ to be highly unlikely.
\bibitem{comment-on-ph-symmetry}
         Since $0 < n < 1$ and $1 < n < 2$ are related
         by particle-hole symmetry, it suffices to
         consider $0 < n < 1$.
\end{thebibliography}
\end{document}